\documentclass[reprint,nofootinbib]{revtex4-2}
\usepackage{amsmath}
\usepackage{amssymb}
\usepackage{amsthm}
\usepackage{bm,bbm}

\usepackage[bookmarks=false]{hyperref}
\hypersetup{colorlinks=true,citecolor=blue,linkcolor=red,%
urlcolor=blue,pdfstartview=FitH,bookmarksopen=true}

\usepackage[T1]{fontenc}
\usepackage[osf,sc]{mathpazo}
\usepackage{dsfont}
\usepackage{overpic}
\usepackage{graphicx}
\usepackage{transparent}
\usepackage{tikz}
\usepackage{tikz-3dplot}
\usetikzlibrary{spy}
\usetikzlibrary{arrows.meta}
\usetikzlibrary{calc}
\usetikzlibrary{decorations.pathreplacing,calligraphy}
\usepackage{intcalc}
\usepackage{physics}
\usepackage[caption=false]{subfig}
\usepackage[shortlabels]{enumitem}
\usepackage{soul}
\usepackage[utf8]{inputenc}
\usepackage{xcolor}
\usepackage{float}
\usepackage{comment}
\usepackage{enumitem}
\usepackage{tcolorbox}

\usepackage{thm-restate}

\newcommand{\cP}{{\cal P}}

\newcommand{\cN}{\mathcal{N}}

\newcommand{\cH}{\mathcal{H}}

\newcommand{\cS}{\mathcal{S}}
\newcommand{\cE}{\mathcal{E}}
\newcommand{\cF}{\mathcal{F}}

\newcommand{\cC}{\mathcal{C}}
\newcommand{\cR}{\mathcal{R}}

\newcommand{\maxover}[1][]{\underset{#1}{\text{max}}}
\newcommand{\minover}[1][]{\underset{#1}{\text{min}}}

\begin{document}
\title{Multipartite entanglement measures: a review}

\author{Mengru Ma}
\affiliation{Key Laboratory of Advanced Optoelectronic Quantum Architecture and Measurement of Ministry of Education, School of Physics, Beijing Institute of Technology, Beijing 100081, China}

\author{Yinfei Li}
\affiliation{Key Laboratory of Advanced Optoelectronic Quantum Architecture and Measurement of Ministry of Education, School of Physics, Beijing Institute of Technology, Beijing 100081, China}

\author{Jiangwei Shang}
\email{jiangwei.shang@bit.edu.cn}
\affiliation{Key Laboratory of Advanced Optoelectronic Quantum Architecture and Measurement of Ministry of Education, School of Physics, Beijing Institute of Technology, Beijing 100081, China}

\date{\today}
\begin{abstract}
Quantum entanglement, a fundamental aspect of quantum mechanics, has captured significant attention in the era of quantum information science. In multipartite quantum systems, entanglement plays a crucial role in facilitating various quantum information processing tasks, such as quantum teleportation and dense coding. In this article, we review the theory of multipartite entanglement measures, with a particular focus on the genuine as well as the operational meaning of multipartite entanglement measures.
By providing a thorough and valuable insight on this field, we hope that this review would inspire and guide researchers in their endeavors to further develop novel approaches for characterizing multipartite entanglement.
\end{abstract}

\maketitle
%

%%%%%%
\section{Introduction}
The concept of entanglement has played an exceptional role in quantum physics ever since its discovery at the beginning of last century~\cite{Schrodinger1935,RevModPhys.81.865}.
Entanglement has been recognized as a key resource in various quantum information processing tasks, such as quantum teleportation~\cite{PhysRevLett.70.1895}, superdense coding~\cite{PhysRevLett.69.2881}, and quantum key distribution~\cite{PhysRevLett.67.661}. Moreover, as a dividing feature between the quantum and classical worlds, how to detect the presence of entanglement has received extensive research interest~\cite{Entdet2009}.
However, describing the structure of quantum entanglement for multipartite systems is still challenging, as there are already an infinite number of entanglement classes for the simple four-qubit system~\cite{PhysRevA.62.062314}.
Recognizing the significance of entanglement as a valuable resource, it becomes natural to explore its quantification through theoretical descriptions. In the seminal work by Vedral \emph{et al.}~\cite{PhysRevLett.78.2275}, the fundamental condition that the amount of entanglement cannot increase under local operations and classical communication (LOCC) is established, laying the foundation for entanglement quantification.
Furthermore, Vidal~\cite{EntanglementMonotones} advanced the field by introducing a family of functions, called entanglement monotones, which are magnitudes that do not increase, on average, under LOCCs.
Examples of entanglement monotones include the entanglement of formation and distillable entanglement~\cite{PhysRevA.54.3824}, relative entropy of entanglement~\cite{PhysRevA.57.1619}, and the robustness of entanglement~\cite{PhysRevA.59.141}.
In addition, entanglement can be quantified using the eigenvalue spectra of the density matrices, which provides a practical means for quantitative entanglement measures~\cite{PhysRevA.54.3824, PhysRevA.57.1619, PhysRevLett.80.2245}.

For the bipartite system, various entanglement measures are designed, such as concurrence~\cite{PhysRevLett.78.5022, PhysRevLett.80.2245}, negativity~\cite{PhysRevA.65.032314}, entanglement of formation~\cite{PhysRevLett.80.2245}, etc.
Since a single Schmidt coefficient governs the two-qubit entanglement, all the entanglement monotones for two-qubit system are equivalent.  Quantifying entanglement for multipartite systems is, however, a great challenge due to the much richer mathematical structures involved as compared to the bipartite case.
The generalization of concurrence, namely the generalized multipartite concurrence, can be found in Ref.~\cite{PhysRevLett.93.230501}.
In Ref.~\cite{PhysRevLett.106.190502}, an entanglement monotone for genuine multipartite entanglement is introduced, which extends the negativity and equals to the negativity in the bipartite case.
Besides, other bipartite entanglement measures such as the relative entropy of entanglement and the robustness of entanglement can be easily generalized to the multipartite scenario.

Though multipartite systems have deeper underlying structures, the definition of multipartite entanglement (ME) can be borrowed straightforwardly from the bipartite case.
Consider an $N$-partite mixed state, the definition of $k$-separability is defined as \cite{gabriel2010criterion}
\begin{equation}
    \rho_{k\text{-SEP}}=\sum_i p_i|\psi_{k\text{-SEP}}^i\rangle\langle\psi_{k\text{-SEP}}^i|\,, \quad \sum_i p_i = 1\,,
\end{equation}
where $p_i$s can be interpreted as a probability distribution, and $|\psi_{k\text{-SEP}}^i\rangle\langle\psi_{k\text{-SEP}}^i|$s are $k$-separable pure states. The state is fully separable iff ${k=N}$ and is called genuine multipartite entanglement (GME) iff it is not biseparable (or ${k=1}$). Being a significant form of entanglement, GME provides notable advantages in various quantum tasks as compared to other types of entanglement. Biseparable states may be less suitable for many applications due to their inherent limitations. Hence, the measures of GME are valuable within the standard LOCC paradigm. For instance, it has been shown that GME is in general essential to establish a multipartite secret key~\cite{PhysRevX.11.041016}.

In recent years, there has been a continuous endeavor to qualitatively and quantitatively characterize the entanglement properties of multipartite systems. This pursuit involves the formulation of entanglement measures through two main approaches: the axiomatic ansatz and the operational approach~\cite{MulEnt}. Under the axiomatic approach, researchers aim to define measures by establishing a set of desired properties that the measures should possess. An essential requirement in this context is entanglement monotonicity~\cite{EntanglementMonotones, 10.1063/1.1495917}.
On the other hand, the operational approach quantifies the utility of a state for specific protocols that rely on entanglement. This perspective evaluates the usefulness of entanglement in achieving certain tasks or performing specific operations. However, computing these measures such as the distillable entanglement tends to be challenging~\cite{PhysRevA.54.3824, Huang2014, PhysRevA.95.062322}.

This review is organized as follows.
In Sec.~\ref{EMOverview}, we first present an overview of entanglement measures, starting from the establishment of the LOCC paradigm to the entanglement monotonicity. We also introduce a few basic axioms that a good entanglement measure should possess. In Sec.~\ref{MEMea}, we explore various common ME measures, and Sec.~\ref{GMEMea} focuses on the GME measures. Subsequently, in Sec.~\ref{tasks}, we delve into the applications of the GME measures in various quantum information tasks and explore the ME measures with an operational meaning. Finally, we provide a summary and outlook in Sec.~\ref{Conclusions}.

%%%%%%
\section{Overview of entanglement measures}
\label{EMOverview}
In this section, we commence by establishing the foundational tenets of quantum entanglement within the scope of the LOCC paradigm. Then, essential guiding postulates that define entanglement measures as well as the convex-roof extension technique will be introduced. We spotlight several basic measures in the end.

\subsection{The LOCC paradigm}
Entanglement is a distinctive quantum property of nonlocality that cannot be found in classical systems. Before presenting the precise mathematical definition of entanglement, we first examine a typical scenario in the field of quantum information science. Imagine that a source prepares a quantum system consisting of two particles. One of the particles is sent to Alice, while the other is sent to Bob, who are located in distant laboratories. Alice and Bob can both apply local operations on their particles, including arbitrary unitary evolution and measurements. We do not pose any restriction on local dimensions in each lab, meaning that Alice, for example, can correlate her particle with an infinite-dimensional system through a unitary evolution. Next, the source may establish a classical correlation between the particles by selecting and sending one from a set of prepared quantum systems based on a classical probability distribution. Finally, Alice and Bob are equipped with quantum memories and classical information channels, enabling them to conduct their local operations with the aid of classical communication.

At this point we have established the paradigm of local operations and classical communication~(LOCC) between Alice and Bob. LOCC is a subset of all the physically realizable operations on quantum states, the significance of which is best celebrated by the quantum teleportation protocol~\cite{PhysRevLett.70.1895}. In quantum teleportation, the source distributes a two-qubit entangled state between Alice and Bob, then Alice performs a joint measurement on the information qubit she wishes to teleport and the shared entangled qubit. Subsequently, Alice sends her measurement outcome to Bob. Upon receiving the outcome, Bob can transform his entangled qubit into an exact copy of Alice's information qubit. More precisely, one entangled two-qubit state and two classical bits are consumed with LOCC operations to transport one qubit.

We notice that entanglement serves as a resource enabling quantum teleportation, and LOCC is the technique used to harness its power. Moreover, it is natural to consider LOCC when discussing entanglement as it does not involve nonlocal quantum operations that may convey quantum information. Indeed, the set of separable or non-entangled states can be considered as the states that result from applying LOCC operations on a pure product state~\cite{PhysRevA.40.4277}. The set of bipartite separable states is therefore defined by
\begin{equation}\label{eq: Def_Ent}
    \rho_\text{bi-SEP} = \sum_i p_i \rho^A_i\otimes \rho^B_i\,, \quad \sum_i p_i = 1\,,
\end{equation}
where $p_i$s refer to a probability distribution and $\rho^X$ is a density matrix supported on the Hilbert space $\mathcal{H}_X$, ${X=\{A,B\}}$. Entangled states refer to states that cannot be expressed in the form of Eq.~\eqref{eq: Def_Ent}.

LOCC is a versatile resource since quantum teleportation implies that an arbitrary physical operation can be performed with LOCC and shared entanglement. Moreover,  LOCC is capable of concentrating entanglement from multiple partially entangled states to a reduced number of more entangled states~\cite{PhysRevA.53.2046}, as well as purifying a noisy entangled state to a pure state~\cite{PhysRevLett.76.722}. However, it is worth noting that there are limitations to the descriptive power of LOCC in capturing nonlocality. Firstly, there are different classes of entangled states that cannot be transformed between each other with LOCC operations only. For bipartite cases, the necessary and sufficient condition that determines the classes is identified by majorization ~\cite{PhysRevLett.83.436}. Next, there are nonlocal states that has no entanglement at all ~\cite{PhysRevA.59.1070}. Even though the preparation of separable states does not require transformation of quantum information, it is possible that a set of mutually orthogonal separable states cannot be faithfully distinguished by LOCC, no matter which one of the states is presented to the local parties.

Finally, we briefly introduce the mathematical definition of LOCC \cite{PhysRevA.54.3824,LOCCChitambar2014}. A general operation, i.e., quantum instrument, on a quantum state is a completely positive trace-preserving~(CPTP) map. For instance, ${\mathcal{A} = \{\mathcal{A}_1,\,\mathcal{A}_2,\,\cdots,\,\mathcal{A}_n\}}$ is a CPTP map if each element $\mathcal{A}_j$ is a completely positive map and $\sum_j{\mathcal{A}_j}$ is trace preserving. Here the maps of interest are bounded linear maps on a set of bounded linear operators acting on a Hilbert space $\mathcal{H}$. Considering a quantum state with $K$ parties, i.e., ${P_1,\,P_2,\,\cdots,\,P_K}$, the Hilbert space is the tensor product of the Hilbert space for each local parties, namely ${\mathcal{H} = \bigotimes_{j = 1}^K \mathcal{H}_{P_j}}$. Henceforth we will use a superscript on an operator or a map to denote the party that it acts on. The one-way local protocol with respect to party $P_k$ is defined by the following operations. Firstly, the local operation $\mathcal{B}$ on party $P_k$ produces a readout $j$, and $P_k$ broadcasts the outcome to all the other parties. Then, the other parties perform local operations $\mathcal{E}_j^{X},\, X\neq P_k$. Thus, a one-way local instrument ${\mathcal{A}^{(P_k)} = \{\mathcal{A}_1,\,\mathcal{A}_2,\,\cdots,\,\mathcal{A}_n\}}$ with respect to party $P_k$ is given by
\begin{equation}
    \mathcal{A}_j = \mathcal{B}_j^{P_k}\otimes \bigotimes_{X\neq P_k} \mathcal{E}_j^{X}\,.
\end{equation}
Extending the discussion to one-round LOCC operations requires the concept of coarse-graining. For two quantum instruments ${\mathcal{A} = \{\mathcal{A}_i|i\in S_\mathcal{A}\}}$ and ${\mathcal{B} = \{\mathcal{B}_i|i\in S_\mathcal{B}\}}$, $\mathcal{A}$ is a coarse-graining of $\mathcal{B}$ if there exists a partition ${S_\mathcal{B} = \cup_i T_{i,\mathcal{B}}}$ of the set of labels $S_\mathcal{B}$, s.t. ${\forall i\in S_\mathcal{A}, \mathcal{A}_i = \sum_{j\in  T_{i,\mathcal{B}}} \mathcal{B}_j}$.
By definition, a quantum instrument $\mathcal{A}$ can be implemented with a one-round LOCC \emph{iff} it is a one-way local operation followed by coarse-graining.

The definition of $r$-round LOCC requires the concept of operation composition ``$\circ$''. An instrument ${\mathcal{A} = \{\mathcal{A}_i|i\in S_\mathcal{A}\}}$ is \emph{LOCC linked} to another instrument $\mathcal{B} = \{\mathcal{B}_i|i\in S_\mathcal{B}\}$ if there exists a set of one-way local instruments ${\bigl\{\mathcal{E}^{(P_j)} = \{\mathcal{E}_{i,j}|i\in S_\mathcal{E}\}, j = 1,2,\cdots\bigr\}}$, such that $\mathcal{A}$ is a coarse-graining of the instrument $\{\mathcal{E}_{i,j}\circ\mathcal{B}_j\}$. This can be interpreted as to first apply the instrument $\mathcal{B}$, then apply a one-round LOCC which depends on the outcome of $\mathcal{B}$. Thus, an instrument is $r$-round LOCC \emph{iff} it is LOCC linked to a $(r-1)$-round LOCC. An immediate observation is that a $(r-1)$-round LOCC is a $r$-round LOCC, while the inverse statement is not always true. For instance, in the task of quantum teleportation between Alice and Bob, one-round LOCC only allows the teleportation of quantum states in a single direction, while two-round LOCC enables the teleportation in both directions.

\subsection{Entanglement monotonicity}
As entanglement is an indispensable resource in quantum information science, an impending task is to quantify it. If a quantum state $\sigma$ can be converted to another state $\rho$ with LOCC, then for local parties equipped with LOCC, any quantum information protocol that requires $\rho$ can also be achieved with $\sigma$. Thus, a fundamental requirement for the quantification of entanglement is that entanglement cannot increase under LOCC operations, i.e., LOCC monotonicity.
An important fact is that for bipartite pure states $\ket{\phi}$ and $\ket{\psi}$, $\ket{\phi}$ can be converted to $\ket{\psi}$ via LOCC \emph{iff} $\ket{\phi}$ is majorized by $\ket{\psi}$, i.e., both states have the same Schmidt dimension $d$, and the Schmidt coefficients $\bigl\{\lambda_i^{(\phi)}\bigr\}$ and $\bigl\{\lambda_i^{(\psi)}\bigr\}$ satisfy
\begin{equation}
    \sum_{i = 1}^k \lambda_i^{(\phi)\downarrow} \le  \sum_{i = 1}^k \lambda_i^{(\psi)\downarrow}\,,\quad \forall k\in\{1,2,\cdots, d\}\,,
\end{equation}
where the superscript $\downarrow$ indicates that the Schmidt coefficients are organized in a descending order. This reveals that some quantum states cannot be related to each other via LOCC. Moreover, two pure states are mutually convertible via LOCC \emph{iff} they can be transformed via local unitary rotations, hence deterministic LOCC is not well suited for the classification of entangled states.

To establish a stronger condition of entanglement ordering, one may consider \textit{stochastic} LOCC~(SLOCC), where $\sigma$ is converted to $\rho$ via LOCC with a non-zero probability \cite{PhysRevA.62.062314}. SLOCC provides a coarse-grained classification of ME, which means that quantum states incomparable via deterministic LOCC can now be related via SLOCC. For two $N$-partite pure states $\ket{\phi}$ and $\ket{\psi}$, they can be transformed to each other via SLOCC or they are SLOCC equivalent \emph{iff} there exists an invertible local operator (ILO) ${\mathcal{L}_1\otimes \mathcal{L}_2\otimes\cdots \otimes \mathcal{L}_N}$, s.t.
\begin{equation}
    \ket{\phi} = \mathcal{L}_1\otimes \mathcal{L}_2\otimes\cdots \otimes \mathcal{L}_N \ket{\psi} \,.
\end{equation}
An SLOCC class can be built from a pure state $\ket{\psi}$ by constructing a convex hull of all pure states that are SLOCC equivalent with $\ket{\psi}$  and of all pure states that can be approximated arbitrarily close
by those SLOCC equivalent pure states \cite{PhysRevLett.87.040401,  Ritz_2019}.

\begin{figure}
	\centering
    \begin{overpic}[width=2.7in]{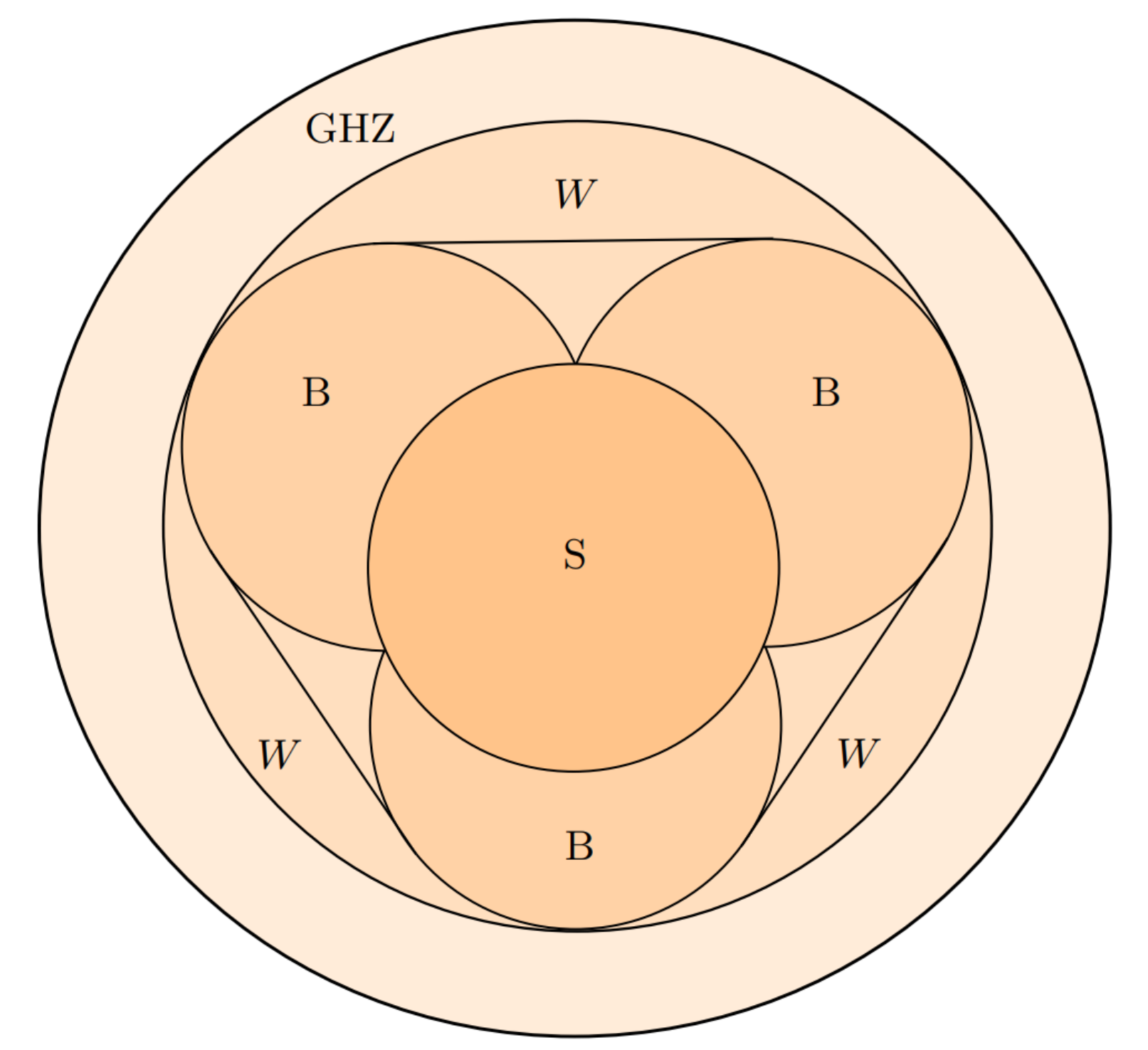}\end{overpic}
\caption{An illustration of the SLOCC classes for three-qubit states~\cite{PhysRevLett.87.040401}. S : the fully-separable class; B: the biseparable class, which is the convex hull of biseparable states w.r.t. an arbitrary partition; $W$: the $W$ class; GHZ: the GHZ class.}
\label{fig:Classes}
\end{figure}

Strong monotonicity of a function $E$ is defined as follows
\begin{equation}
    \sum_i{p_i E\bigl(\sigma_i\bigr)}\le E(\rho)\,,
\end{equation}
where $\{p_i, \sigma_i\}$ is an ensemble produced by an arbitrary LOCC channel $\Lambda_{\text{LOCC}}$ acting on the multipartite quantum state $\rho$, such that
\begin{equation}
    \Lambda_{\text{LOCC}}(\rho)=\sum_i{p_i\sigma_i}\,.
\end{equation}
For states in the same SLOCC class, the monotonicity provides the entanglement ordering. However, there are already two types of genuinely entangled states for three-qubit systems~\cite{PhysRevA.62.062314,SDThr_PhysRevLett.85.1560}, as is shown in Fig.~\ref{fig:Classes}, resulting in the fact that different entanglement measures may lead to distinct entanglement ordering for multipartite quantum states.

\subsection{Postulates of entanglement measures}
An entanglement measure $E(\rho)$ should have some desirable properties, while not all of the following properties
are fulfilled by all the discussed entanglement quantifiers.
\begin{enumerate}[(1)]
\item Entanglement vanishes for separable states. If $\rho$ is separable, then $E(\rho)=0$.
\item $E(\rho)$ cannot be increased by LOCC. For any LOCC operation $\Lambda_{\text{LOCC}}$,
\begin{equation}\label{EMp2_1}
  E\bigl(\Lambda_{\text{LOCC}}(\rho)\bigr)\leq E(\rho)\,.
\end{equation}
And there is a stronger condition, if the ensemble $\{p_k, \rho_k\}$ is obtained from the state $\rho$ by applying LOCC operations, then the measure should not increase on average
\begin{equation}\label{EMp2_2}
  \sum_kp_kE(\rho_k)\leq E(\rho)\,.
\end{equation}
\item Convexity. Most entanglement measures are convex, that is, the entanglement measure cannot increase under convex combination of two or more states,
\begin{equation}\label{EMp4}
        E\Bigl(\sum_k p_k \rho_k\Bigr) \leq\sum_kp_kE(\rho_k)\,.
\end{equation}
\item Additivity. If Alice and Bob share $n$ copies of the same state $\rho$, then
\begin{equation}\label{EMp5}
      E\bigl(\rho^{\otimes n}\bigr)=nE(\rho)\,.
\end{equation}
    If there are two different states, then
    \begin{equation}\label{EMp52}
    E\bigl(\rho_1\otimes\rho_2\bigr)=E\bigl(\rho_1\bigr)+E\bigl(\rho_2\bigr)\,.
    \end{equation}
And if the measure satisfies
\begin{equation}\label{EMp5_2}
      E\bigl(\rho^{\otimes n}\bigr)\leq nE(\rho)\,,
\end{equation}
it is subadditive.

\item Asymptotic continuity. For states $\rho_n$ and $\sigma_n$ acting on the Hilbert space ${\cH_n=\cH_n^A\otimes \cH_n^B, n\in \mathbb{N}}$, we have \cite{EntanglementMonotones, PhysRevLett.84.2014, 10.1063/1.1495917}
    \begin{equation}\label{Continuity}
      \|\rho_n-\sigma_n\|_1\to 0 \Rightarrow \frac{|E(\rho_n)-E(\sigma_n)|}{\log_2 {\text{dim}\cH_n}}\to 0\,,
    \end{equation}
    where $\|\cdot\|_1$ is the trace norm. The measures that satisfy the postulate of asymptotic continuity are valuable for estimating the distillable entanglement, which will be discussed in Sec.~\ref{DE and EC}.
\end{enumerate}

Postulate (1) and Eq.~\eqref{EMp2_1} in postulate (2) constitute the essential criteria that any entanglement measure must satisfy. An entanglement measure is also known as an entanglement monotone \cite{EntanglementMonotones} if it satisfies postulates (1), (3), and Eq.~\eqref{EMp2_2} in postulate (2).
Therefore, an entanglement monotone is an entanglement measure, but not vice versa.

\subsection{Convex-roof extension}
The construction of convex roof is commonly used to quantify the entanglement of mixed states~\cite{EntanglementMonotones}. Given an entanglement measure of a pure state $E(\ket{\phi})$, it can be extended to a mixed state by means of convex roof, such that
\begin{equation}\label{convexroof}
  E(\rho)=\inf_{\{p_k,\ket{\phi_k}\}}\sum_kp_k E\bigl(\ket{\phi_k}\bigr)\,,
\end{equation}
where the infimum is taken over all ensembles $\bigl\{p_k,\ket{\phi_k}\bigr\}$ of ${\rho=\sum_kp_k\ket{\phi_k}\!\bra{\phi_k}}$. The optimal ensemble that achieves this infimum is called the optimal ensemble for $E$. Although concurrence of  the two-qubit mixed states can be obtained analytically, typically the convex roof cannot be easily determined. However, there are a number of effective schemes to compute the convex roof, such as providing a lower bound of it as an approximation.

The convex-roof construction is a reliable method for quantifying entanglement. It produces an entanglement measure that is guaranteed to be convex. Moreover, if a measure is monotonic on pure states, then its convex-roof extension is also monotonic on mixed states~\cite{RevModPhys.81.865}. Thus, to verify the monotonicity of a convex roof entanglement measure, one only needs to examine the monotonicity for pure states. It is worth noting that the properties of the convex roof can be translated to ones for the concave roof.

%%%%%%
\subsection{Entanglement of formation}
For pure state ${\ket{\psi}\in\cH_{AB}}$, the entanglement of formation (EoF) $E_f$ is defined by~\cite{PhysRevA.54.3824,PhysRevLett.80.2245}
\begin{equation}
    E_f(\ket{\psi}):=S(\rho_A)\,,
\end{equation}
where ${S(X)=-\Tr (X \log_2 X)}$ is the von Neumann entropy of the system $X$, ${\rho_A=\Tr_B\bigl(\ket{\psi}\!\bra{\psi}\bigr)}$. For mixed states, the EoF  is defined as
\begin{equation}
    E_f(\rho):=\minover[\{p_i,\ket{\psi_i}\}]\sum_ip_iE_f(\ket{\psi_i})\,,
\end{equation}
where the minimum is taken over all pure states that decompose $\rho$. In general EoF lacks the property of additivity, but a class of states the EoF of which is indeed additive is introduced in Ref.~\cite{PhysRevA.101.032301}.

\subsection{Concurrence}
Concurrence is one of the most celebrated entanglement measures, which is the first entanglement measure built with the convex-roof extension. In two-qubit cases, it has been proven that as the entanglement of formation increases, concurrence monotonically decreases
\cite{PhysRevLett.80.2245}.
Moreover, it has been successfully extended to higher-dimensional bipartite systems as
\begin{equation}\label{eq:bipartiteconc}
    \mathcal{C}(\ket{\psi}) = \sqrt{1-\Tr\bigl(\rho^2\bigr)}\,,
\end{equation}
where $\rho$ is the reduced density matrix of the subsystem \cite{PhysRevA.64.042315}.
The generalized bipartite concurrence Eq.~\eqref{eq:bipartiteconc} is also an entanglement monotone under SLOCC~\cite{PhysRevA.74.052303}.
For a bipartition $\{A|B\}$ of a pure state $\ket{\psi}_{AB}$, we denote the corresponding concurrence as $\mathcal{C}(|\psi\rangle_{A|B})$ or $\mathcal{C}_{A|B}$.

\subsection{Robustness of entanglement}
For a given entangled state $\rho$ and a separable state $\rho_s$, the robustness of entanglement $R_s(\rho)$ \cite{PhysRevA.59.141} is defined as the minimal $t\ge 0$ such that the state
\begin{equation}
    \frac{1}{1+t}(\rho+t\rho_s)
\end{equation}
is separable. The definition can be extended to the generalized robustness of entanglement case \cite{PhysRevA.67.054305}, where $\rho_s$ can be any density matrix. In general, the generalized robustness of entanglement quantifies the minimal amount of a general state that needs to be mixed with $\rho$ so that the mixture is separable.

It is known that the exact computation of entanglement measures tends to be unfeasible. Consequently, one can establish methods to connect the violation of entanglement witnesses
to lower bounds for the generalized robustness \cite{PhysRevLett.127.010401}.
For a quantum state, the variance of a self-adjoint operator $O_k$ is ${\min_{s_k}\langle (O_k - s_k\openone)^2\rangle}$, where $s_k$ is a real number. Let ${\mathbf{s} = \{s_1,s_2,\dots\}}$, then the general robustness is lower bounded by $- \min_{
\mathbf{s}}\langle W(\mathbf{s})\rangle/m$, where
\begin{equation}
    W(\mathbf{s}) = \sum_k (O_k - s_k\openone)^2 - B\,,
\end{equation}
$B$ is a self-adjoint operator, $m = \sum_{k} \lambda_{\max} (O_k)^2 - \lambda_{\min}(B)$ with $\lambda_{\min}$ and $\lambda_{\max}$ denoting the minimal and maximal eigenvalues.

%%%%%%
\section{ME measures}\label{MEMea}
In this section we present examples of ME measures and discuss how some of them can be defined from measures of the bipartite case. In addition, we introduce a new class of ME measures, namely the complete ME measure~(CMEM)~\cite{PhysRevA.101.032301}.

Let $\cS_X$ be the set of density operators
acting on the Hilbert
space $\cH_X$, a function ${E^{(m)}:\cS_{A_1A_2\cdots A_m}\to\mathbb{R}_+}$ is called an $m$-partite entanglement measure if it satisfies~\cite{RevModPhys.81.865,PhysRevA.83.062325}:
\begin{enumerate}[(E1)]
\item $E^{(m)}(\rho)=0$ if $\rho$ is fully separable;
\item $E^{(m)}$ cannot increase under $m$-partite LOCC.
\end{enumerate}

\subsection{Squashed entanglement}
Many of the axiomatic measures of bipartite states can be extended to the multipartite case.
For a given state $\rho_{AB}$, the squashed entanglement is given by
\begin{equation}\label{SquEnt}
  E_{sq}(\rho_{AB})=\frac{1}{2}\inf I\bigl(A:B|E\bigr)\,,
\end{equation}
where the infimum is taken over all density matrices $\rho_{ABE}$ such that ${\Tr_E\rho_{ABE}=\rho_{AB}}$, and the conditional mutual information is defined as
\begin{equation}\label{ConMI}
  I\bigl(A:B|E\bigr)=S(AE)+S(BE)-S(ABE)-S(E)\,.
\end{equation}
 Multipartite $c$-squashed entanglement is defined as follows \cite{5075874}. For the $m$-party state $\rho_{A_1, A_2, \cdots, A_m}$,
\begin{equation}\label{cSquEnt}
  E_{sq}^c(\rho_{A_1, A_2, \cdots, A_m})=\inf I\bigl(A_1:A_2:\cdots:A_m|E\bigr)\,,
\end{equation}
where the infimum is taken over the extension states ${\sigma_{A_1, A_2, \cdots,A_m,E}}$ of the form ${\sum p_i\rho^i_{A_1, A_2, \cdots, A_m}\otimes\ket{i}_E\!\bra{i}}$. The conditional mutual information is a sum of conditional bipartite mutual information
\begin{align}\label{MConMI}
  I&(A_1:A_2:\cdots:A_m|E)=I\bigl(A_1:A_2|E\bigr)+\\
  &\,\,\,I\bigl(A_3:A_1A_2|E\bigr)+\cdots+I\bigl(A_m:A_1\cdots A_{m-1}|E\bigr)\,.\nonumber
\end{align}
Thus $E_{sq}^c$ is an entanglement measure which is obtained through the mixed convex roof of the multipartite quantum mutual information function, and it is indeed a good entanglement measure. In Ref.~\cite{5075874}, the authors also define the multipartite $q$-squashed entanglement as a generalization of the bipartite squashed entanglement. For the $m$-party state $\rho_{A_1, A_2, \cdots, A_m}$,
\begin{equation}\label{qSquEnt}
  E_{sq}^q(\rho_{A_1, A_2, \cdots, A_m})=\inf I\bigl(A_1:A_2:\cdots:A_m|E\bigr)\,,
\end{equation}
where the infimum is taken over the extension states $\sigma_{A_1, A_2, \cdots,A_m,E}$ of $\rho_{A_1, A_2, \cdots, A_m}$. It is proved that the $q$-squashed entanglement satisfies several important properties, including the monotonicity under LOCC, convexity, additivity, and asymptotic continuity. Furthermore, it is shown that multipartite $q$-squashed entanglement is an upper bound on multipartite distillable key; see Sec.~\ref{DistillableKey}.

\subsection{Three-tangle}
The three-tangle (or residual entanglement) $\tau$ is an entanglement measure for three-qubit states \cite{PhysRevA.61.052306}. It is defined as follows
\begin{equation}\label{3tangle}
  \tau\equiv\tau_{ABC}=\cC^2_{A|BC}-\cC^2_{AB}-\cC^2_{AC}\,,
\end{equation}
where $\cC^2_{A|BC}$ is the squared concurrence between $A$ and the pair $BC$, $\cC^2_{AB}$ is the squared concurrence between $A$ and $B$, and $\cC^2_{AC}$ is the squared concurrence between $A$ and $C$. The three-tangle is permutationally invariant of the parties.
% $\sqrt{\tau} is also monotone$
It is noted that bipartite measures cannot determine the inequivalence of the GHZ and $W$ classes, but three-tangle can. $\tau(\psi_{\textrm{GHZ}})\ne 0$ for any state in the GHZ class, while it vanishes for any state in the $W$ class.

The generalization of the three-tangle for $N$ qubits, namely, the $N$-tangle is discussed in Ref.~\cite{PhysRevA.63.044301}.

\subsection{Global entanglement}
There exist measures that can be calculated as straightforward combinations of bipartite entanglement measures. Global entanglement of tripartite system, for instance, is the sum of squared concurrences between a single qubit versus all other qubits \cite{10.1063/1.1497700, Brennen2003},
\begin{equation}\label{GloEnt}
  Q=\cC^2\bigl(\ket{\phi}_{A|BC}\bigr)+ \cC^2\bigl(\ket{\phi}_{B|AC}\bigr)+\cC^2\bigl(\ket{\phi}_{C|AB}\bigr)\,.
\end{equation}
Its monotonicity under LOCC is simply inherited from the bipartite measures.

By using the relations \cite{PhysRevA.61.052306}
\begin{equation}\label{resicon}
\cC^2(\ket{\phi}_{i|jk})=\tau_{ABC}+\cC^2(\rho_{ij})+\cC^2(\rho_{ik})\,,
\end{equation}
where ${\rho_{ij}=\Tr_{k}\bigl(\ket{\phi}_{ijk}\!\bra{\phi}\bigr)}$ and $i, j, k$ are distinct systems in $\{A,B,C\}$.
Then the global entanglement can be written as
\begin{equation}\label{peritangle}
Q=3\tau_{ABC}+2\bigl[\cC^2(\rho_{AB})+\cC^2(\rho_{AC})+\cC^2(\rho_{BC})\bigr]\,.
\end{equation}

\subsection{Schmidt measure}
For a pure state ${\ket{\phi}\in\cH, \cH=\mathbb{C}^{d_1}\otimes\mathbb{C}^{d_2}\otimes\cdots\otimes\mathbb{C}^{d_N}}$ in the composite system with parties ${A_1,A_2,\cdots, A_N}$, the Schmidt measure is defined as \cite{PhysRevA.64.022306}
\begin{equation}\label{SchMeaPu}
  E_P(\ket{\phi}\!\bra{\phi})=\log_2r\,,
\end{equation}
where
\begin{equation}\label{dePure}
  \ket{\phi}=\sum_{i=1}^R\alpha_i\ket{\phi_{A_1}^{(i)}}
  \otimes\ket{\phi_{A_2}^{(i)}}\otimes\cdots\otimes\ket{\phi_{A_N}^{(i)}}\,,
\end{equation}
and ${\ket{\phi_{A_j}^{(i)}}\in \mathbb{C}^{d_j}, j=1,2,\cdots,N, \alpha_i\in\mathbb{C}}$, and $r$ is the minimal number of product terms $R$ in the decomposition of $\ket{\phi}$. For a bipartite system with parties $A_1$ and $A_2$, the minimal number of product terms $r$ is given by the Schmidt rank of the state. One can extend the definition to the full state space $\cS(\cH)$ by using a convex-roof construction. For $\rho\in\cS(\cH)$,
\begin{equation}\label{SchMeaMix}
  E_P(\rho)=\inf\sum_i\lambda_iE_P(\ket{\phi_i}\!\bra{\phi_i})\,,
\end{equation}
where the infimum is taken over all possible convex combinations of the form ${\rho=\sum_i\lambda_i\ket{\phi_i}\!\bra{\phi_i}}$ with ${0\le \lambda_i\le 1}$ for all $i$.

The measure is zero iff the state is fully separable. Therefore, it cannot distinguish GME from bipartite entanglement.

\subsection{Complete ME measure}
In Ref.~\cite{PhysRevA.101.032301}, it is shown that a \emph{unified} ME measure should
also satisfy
\begin{enumerate}[(E3)]
\item the unification condition, i.e., $E^{(3)}$ is consistent with $E^{(2)}$.
\end{enumerate}
Also, a CMEM not only satisfies the conditions (E1)-(E3), but additionally satisfies
\begin{enumerate}[(E4)]
\item the hierarchy condition, i.e.,
\begin{equation}
\begin{aligned}
        &\forall \rho_{ABC},\,{X,Y,Z\in\{A,B,C\},}\\
    &E^{(3)}(\rho_{ABC})\ge E^{(2)}(\rho_{X|YZ})\ge E^{(3-2)}(\rho_{ABC})\,,
\end{aligned}
\end{equation}
where
\begin{equation}
\begin{aligned}
    E^{(3-2)}(\ket{\psi}):=\min\{&E^{(2)}(\ket{\psi}_{A|BC}),E^{(2)}(\ket{\psi}_{B|AC}),\,\\
    &\,E^{(2)}(\ket{\psi}_{C|AB})\}\,.
\end{aligned}
\end{equation}
\end{enumerate}
One can find that the multipartite extension of EoF, tripartite concurrence and tripartite tangle proposed in Ref.~\cite{PhysRevA.101.032301} are CMEMs. Moreover, these multipartite extensions are tightly completely monogamous, that is, they satisfy
\begin{equation}
    E^{\alpha}(\rho_{ABC})\ge E^{\alpha}(\rho_{A|BC})+E^{\alpha}(\rho_{BC})
\end{equation}
for some $0<\alpha<\infty$. Here the superscripts of multipartite entanglement measures $E^{(2,3)}$ are omitted.

%%%%%%
\section{GME measures}
\label{GMEMea}
Numerous entanglement measures, particularly for the bipartite case, have been proposed.
Yet, the situation is much more complicated for multipartite systems. How to quantify genuine
entanglement contained in multipartite quantum states is still lacking in research.

If an $N$-partite state is not biseparable then it is called genuinely $N$-partite entangled.
A measure of GME $E(\rho)$ should at least satisfy~\cite{PhysRevLett.127.040403}
\begin{enumerate}[(1)]
\item $E(\rho)=0, \forall \rho\in \cS_{\text{bi-SEP}}$, where $\cS_{\text{bi-SEP}}$ is the set of biseparable states;
\item $E(\rho)>0, \forall \rho\in \cS_{\text{GME}}$, where $\cS_{\text{GME}}$ is the set of all GME states;
\item (Strong monotonicity) ${E(\rho)\ge \sum_{i}p_iE(\rho_i)}$, where $\{p_i, \rho_i\}$ is an ensemble of states produced by applying LOCC operations on $\rho$;
\item $E\bigl(U_\text{local}\rho U^{\dagger}_\text{local}\bigr)=E(\rho)$, where $U^{\dagger}_\text{local}$ is an arbitrary local unitary operator.
\end{enumerate}

\subsection{GME-concurrence}
For an $N$-partite pure state $\ket{\phi}\in \cH_1\otimes \cH_2\otimes\cdots\otimes\cH_N$,
 where dim${(\cH_i)=d_i, i=1,2,\cdots, N}$, the GME-concurrence is defined as \cite{PhysRevA.83.062325}
\begin{equation}\label{GMECP}
  \cC_{\text{GME}}(\ket{\phi}):=\minover[\gamma_i\in \gamma]\sqrt{2\Bigl[1-\Tr(\rho_{A_{\gamma_i}}^2)\Bigr]}\,,
\end{equation}
where $\gamma=\{\gamma_i\}$ represents the set of all possible bipartitions $\{A_i|B_i\}$
of $\{1, 2, \cdots, N\}$. The GME-concurrence can be generalized to mixed states $\rho$ via a convex-roof construction, i.e.,
\begin{equation}\label{GMECM}
  \cC_{\text{GME}}(\rho)=\inf_{\{p_i,\ket{\phi_i}\}}\sum_{i}p_i\cC_{\text{GME}}\bigl(\ket{\phi_i}\bigr)\,,
\end{equation}
where the infimum is taken over all possible decompositions ${\rho=\sum_ip_i\ket{\phi_i}\!\bra{\phi_i}}$.
It is a special case of $k$-ME concurrence when ${k=2}$. The measure $k$-ME concurrence can detect all $k$-nonseparable states, and exhibit properties such as being an entanglement monotone, vanishing on $k$-separable states, convexity, subadditivity, invariant under local unitary transformations \cite{PhysRevA.86.062323}.
The lower bounds for GME-concurrence are introduced in Refs.~\cite{PhysRevA.83.062325,PhysRevA.86.062323}.

\subsection{Geometric measure of entanglement}
Motivated by the geometric measure (GM) of entanglement introduced by Wei and Goldbart \cite{PhysRevA.68.042307}, Ref.~\cite{PhysRevA.81.012308} presents a GME
measure called the generalized geometric measure (GGM). Consider an $N$-partite pure quantum state $\ket{\phi_N}$, and let
\begin{equation}\label{MaOvlap}
  g_{\text{max}}(\phi_N)=\maxover[\psi]|\braket{\psi}{\phi_N}|\,,
\end{equation}
where the maximum is over all pure quantum states $\ket{\psi}$ that are not genuinely $N$-party entangled. And $g_{\text{max}}$ quantifies the closeness of the state $\ket{\phi_N}$ to all pure quantum states that are not genuinely multipartite entangled. A pure quantum state of $N$ parties is considered to be genuinely $N$-party entangled if it cannot be expressed as a product state of any bipartite partition. Then the GGM is defined as
\begin{equation}\label{GGM}
  E_G(\ket{\phi_N})=1-g^2_{\text{max}}(\ket{\phi_N})\,.
\end{equation}
It is noted that the maximization in $g_{\text{max}}$ is only over pure states that are a product over all bipartite partitions. The GGM $E_G$ is vanishing for all pure states that are not genuine multipartite entangled and non-vanishing for others. The measure is computable for a multipartite pure state of an arbitrary
number of parties, and of arbitrary dimensions, and it is monotonically decreasing under LOCC.

The generalized geometric measure can be
written in a computable form for all multipartite pure quantum states as \cite{PhysRevA.81.012308,PhysRevA.86.052337}
\begin{align}\label{GGMComputable}
  E_G(\phi_N)=1-\max\Bigl\{\lambda_{A|B}^2|&A\cup B=\{1,2,\cdots, N\},\nonumber\\
  &A\cap B=\varnothing\Bigr\}\,,
\end{align}
where $\lambda_{A|B}$ is the maximal Schmidt coefficient of $\ket{\phi_N}$ in the bipartite split ${A|B}$. For the three-qubit pure states $\ket{\phi}$, the GGM reduces to
\begin{equation}\label{3qGGMComputable}
  E_G=1-\max\bigl\{\lambda_A^2, \lambda_B^2,\lambda_C^2\bigr\}\,,
\end{equation}
where $\lambda_A^2$ is the maximal eigenvalue of ${\rho_A=\Tr_{BC}\bigl(\ket{\phi}\!\bra{\phi}\bigr)}$, similarly for $\lambda _B^2$ and $\lambda_C^2$.

\subsection{Tripartite negativity}
An ideal measure of the full tripartite entanglement of
three qubits should have at least the following characteristics:
i) to be zero for any fully separable or biseparable state and non-zero for any fully entangled state; ii) to be invariant under local unitary (LU); iii) to be nonincreasing under LOCC.
Tripartite negativity is defined according to these three conditions, which is expressed as \cite{sabin2008classification}
\begin{equation}\label{TriNeg}
  \cN_{ABC}=\bigl(\cN_{A|BC}\cN_{B|AC}\cN_{C|AB}\bigr)^{\frac{1}{3}}\,,
\end{equation}
where the bipartite negativities are defined as
\begin{equation}\label{biNeg}
  \cN_{I|JK}=-2\sum_i\varepsilon_i\bigl((\rho^I)^T\bigr)\,,
\end{equation}
where $\varepsilon_i\bigl((\rho^I)^T\bigr)$ denotes the negative eigenvalues of $(\rho^I)^T$, and $(\rho^I)^T$ is the partial transpose of $\rho$ with respect to subsystem $I$, namely, $\langle i_I, j_{JK}|(\rho^I)^T|k_I, l_{JK}\rangle=\langle k_I, j_{JK}|\rho|i_I,l_{JK}\rangle$, with $I=A,B,C$ and $JK=BC,AC,AB$, respectively. The tripartite negativity fulfills the above three conditions for pure states.

\subsection{Concurrence fill}
The challenge of establishing a proper entanglement ordering for genuine three-qubit entanglement is first tackled by the proposal of concurrence fill~\cite{PhysRevLett.127.040403,Xie2021ContPhy}.
For an arbitrary pure three-qubit state $\ket{\phi}_{ABC}$ shared by three parties $A, B, C$, the concurrence between the bipartition $i$ and $jk$ is ${\cC\bigl(\ket{\phi}_{i|jk}\bigr)=\sqrt{2\bigl[1-\Tr(\rho_i^2)\bigr]}}$, with ${\rho_i=\Tr_{jk}\bigl(\ket{\phi}_{ijk}\!\bra{\phi}\bigr)}$ and $i, j, k$ are distinct systems in $\{A,B,C\}$. The three one-to-other bipartite entanglements are not independent, and they follow the relation \cite{PhysRevA.92.062345}
\begin{equation}
\cC^2\bigl(\ket{\phi}_{i|jk}\bigr)\le \cC^2\bigl(\ket{\phi}_{j|ik})+\cC^2(\ket{\phi}_{k|ij}\bigr)\,.
\end{equation}
These three squared bipartite concurrences can be geometrically interpreted as
the lengths of three sides of a triangle, which is the so-called concurrence triangle.
Then the concurrence fill is defined as the square-root of the area of the concurrence triangle \cite{PhysRevLett.127.040403}
\begin{align}\label{Confill}
  \cF&_{123} =  \Bigl[\frac{16\cP_{ABC}}{3}\bigl(\cP_{ABC}-\cC^2(\ket{\phi}_{A|BC})\bigr) \\
                          & \times \bigl(\cP_{ABC}-\cC^2(\ket{\phi}_{B|AC})\bigr)\bigl(\cP_{ABC}-\cC^2(\ket{\phi}_{C|AB})\bigr)\Bigr]^{1/4}\,,\nonumber
\end{align}
with half-perimeter
\begin{equation}\label{Peri}
  \cP_{ABC}=\frac{1}{2}\Bigl[\cC^2\bigl(\ket{\phi}_{A|BC}\bigr)+\cC^2\bigl(\ket{\phi}_{B|AC}\bigr)+\cC^2\bigl(\ket{\phi}_{C|AB}\bigr)\Bigr]\,.
\end{equation}

The concurrence fill was studied in the experiments of three-flavor neutrino oscillations~\cite{Li2022exp}.
The extension of concurrence fill to four-partite systems, inspired by a richer geometrical interpretation, is presented in \cite{mishra2022geometric}, which interprets as a combination of areas of cyclic quadrilateral and triangle structures, resulting from two types of bipartition. In addition to the proposed three-qubit concurrence triangle, the authors in Ref.~\cite{Guo_2022} show that the
triangle relation holds for any continuous entanglement measures and systems of any dimension. They also present tripartite entanglement
measures
\begin{align}
    \cE_{g-123}&=\delta(xyz)(x+y+z)\,,\\
    \cE_{123}&=x+y+z\,,
\end{align}
with bipartite entanglement measure $E_{A|BC}=x, E_{AB|C}=y, E_{B|AC}=z$, and ${\delta(x)=0}$ whenever $x=0$, otherwise ${\delta(x)=1}$. It is shown that
the convex-roof extension of ${\cE_{g-123}}$ is a genuine tripartite entanglement measure.  Using similar argument, four-partite entanglement measures can be constructed. Moreover, they establish four-partite entanglement measures via the tripartite entanglement measures.

The concurrence fill is proposed for GME quantification and satisfies the property of faithfulness and smoothness, but it is later proven that the concurrence fill can be increased under some LOCC operations~\cite{PhysRevA.107.032405}. Nevertheless, numerical evidence supports that if the edges of the triangle are chosen as bipartite concurrence, then the area is nonincreasing under LOCC. Subsequently, various GME measures are proposed, which are based on the bipartite entanglement measures~\cite{PhysRevResearch.4.023059,https://doi.org/10.1002/andp.202300305,PhysRevA.107.032405,PhysRevA.107.012409,JIN2023106155}. For any tripartite pure state $\ket{\phi}_{ABC}$, Ref.~\cite{JIN2023106155} defines the genuine tripartite entanglement measure as
\begin{align}
  \cF&_{3} =  \Bigl[\frac{16\cP_{3}}{3}\bigl(\cP_{3}-\cC(\ket{\phi}_{A|BC})\bigr) \\
                          & \times \bigl(\cP_{3}-\cC(\ket{\phi}_{B|AC})\bigr)\bigl(\cP_{3}-\cC(\ket{\phi}_{C|AB})\bigr)\Bigr]^{1/2}\,,\nonumber
\end{align}
with half-perimeter
\begin{equation}
  \cP_{3}=\frac{1}{2}\Bigl[\cC\bigl(\ket{\phi}_{A|BC}\bigr)+\cC\bigl(\ket{\phi}_{B|AC}\bigr)+\cC\bigl(\ket{\phi}_{C|AB}\bigr)\Bigr]\,.
\end{equation}
Here, three concurrences are used to represent the lengths
of the three edges of a concurrence triangle.

\subsection{Geometric mean of bipartite concurrences}
The geometric mean of bipartite concurrences (GBC) $\mathcal{G}$ satisfies the figure of merits of concurrence fill and qualifies as an entanglement monotone~\cite{PhysRevResearch.4.023059}, while lacks a geometric interpretation.
Denote the set of all possible bipartitions as $\alpha$ and its cardinality ${c(\alpha)}$, we have
\begin{equation}
    \mathcal{G}(\rho)=\sqrt[{c(\alpha)}]{\prod\limits_{\alpha_j\in\alpha}\cC_{A_{\alpha_j}B_{\alpha_j}}}\,.
\end{equation}
The strong monotonicity of GBC is proved as follows.
Bipartite concurrence satisfies the property of strong monotonicity~\cite{PhysRevA.74.052303}, i.e.,
\begin{equation}\label{eq35}
    \sum_i{p_i\cC_{AB}\bigl(\sigma_i\bigr)}\le\cC_{AB}(\rho)\,,
\end{equation}
where $\{p_i, \sigma_i\}$ is an ensemble produced by an arbitrary LOCC channel $\Lambda_{\text{LOCC}}$ acting on the multipartite quantum state $\rho$, such that
\begin{equation}
    \Lambda_{\text{LOCC}}(\rho)=\sum_i{p_i\sigma_i}\,.
\end{equation}
Note that $\Lambda_{\text{LOCC}}$ is also an LOCC channel with respect to the bipartition $\{A|B\}$.
Hereby we examine the monotonicity for pure states, i.e., when $\rho$ and $\sigma_i$ are all pure states. Since all the bipartite concurrences in the definition of GBC satisfy Eq.~\eqref{eq35}, we have
\begin{equation}
\begin{aligned}
    \mathcal{G}(\rho)\ge&\sqrt[{c(\alpha)}]{\prod\limits_{\alpha_j\in\alpha}\Bigl(\sum_i{p_i\cC_{A_{\alpha_j}B_{\alpha_j}}\bigl(\sigma_i\bigr)\Bigr)}}\\
    \ge&\sum_i{p_i\Biggl(\sqrt[{c(\alpha)}]{\prod\limits_{\alpha_j\in\alpha}\mathcal{C}_{A_{\alpha_j}B_{\alpha_j}}\bigl(\sigma_i\bigr)}\Biggr)}=\sum_i{p_i\mathcal{G}\bigl(\sigma_i\bigr)}\,,
\end{aligned}
\end{equation}
where the concavity of the geometric mean function defined on the convex set with all the variables ranging from 0 to 1 is employed, which can be easily proven with the Mahler's inequality.
In fact, any multivariable concave function of bipartite entanglement monotones preserve the monotonicity.

GBC is the product of bipartite entanglement monotones with an exponential order $c(\alpha)$ specified by the number of possible bipartitions. However, the exponential order can be further improved. The square root of the product of all the bipartite concurrences of a three-qubit pure state is also proven to be an entanglement monotone \cite{PhysRevA.107.032405}, which improves the exponential order in GBC from $1/3$ to $1/2$. Also, it can be generalized to any pure $N$-qubit state as the product of all the bipartite concurrences to the order of $1/2^{N-2}$.  Moreover, it is found that the negativity is exactly equal to the geometric mean
of bipartite concurrences for the three-qubit pure states, although the negativity is always less than or equal to
the latter for three-qubit mixed states~\cite{PhysRevA.107.052403}.

\subsection{GME measures based on teleportation capability}
In order to quantify GME as a resource, one needs to consider the usefulness of GME in quantum information processing tasks. For three-qubit teleportation, the GHZ state results in better teleportation fidelity than the $W$ state. Consequently, many existing GME measures, such as the concurrence fill, are \textit{proper} in the sense that they are larger for GHZ state than $W$ state. In Ref.~\cite{Choi2023_10.1038/s41598-023-42052-x}, the authors extended this idea by considering the teleportation capability for general pure states.

The three-qubit teleportation scheme with a resource state $|\psi\rangle_{ABC}$ can be described as follows: first measure a system $k$ and observe the outcome, then perform two-qubit teleportation on the remaining systems. When the teleportation is performed with qubits $i$ and $j$, the maximal average fidelity of teleportation is \cite{PhysRevA.72.024302}
\begin{equation}
    F_{ij}(|\psi\rangle_{ABC})=\frac{\sqrt{\tau_{ABC}+\cC_{ij}^2}+2}{3},
\end{equation}
where $\cC_{ij}$ is the concurrence for the reduced density operator $\rho_{ij}$ of $|\psi\rangle_{ABC}$, and $i,j$ are distinct systems in $\{A,B,C\}$.
 Also, let $\mathcal{T}_{ij}(|\psi\rangle_{ABC}) = 3F_{ij}(|\psi\rangle_{ABC}) - 2$. The authors define GME measures $\mathcal{T}_\text{min}$ and $\mathcal{T}_\text{GM}$ as
\begin{align}
    \mathcal{T}_\text{min} & \equiv \min\{\mathcal{T}_{ij},\,\mathcal{T}_{jk},\,\mathcal{T}_{ik}\}\,,\\
    \mathcal{T}_\text{GM} & \equiv \sqrt[3]{\mathcal{T}_{ij} \mathcal{T}_{jk} \mathcal{T}_{ik}}\,,
\end{align}
respectively.
Both $\mathcal{T}_\text{min}$ and $\mathcal{T}_\text{GM}$ satisfy the requirements for GME measures.
Moreover, the authors prove that if both $\mathcal{T}_{ij}$ and $ \mathcal{T}_{ik}$ are strictly positive, then so does $\mathcal{T}_{jk}$. Thus, they define two additional GME measures
\begin{align}
    \mathcal{T}_\text{min}^{(i)} & \equiv \min\{\mathcal{T}_{ij},\, \mathcal{T}_{ik}\}\,,\\
    \mathcal{T}_\text{GM}^{(i)} & \equiv \sqrt{\mathcal{T}_{ij} \mathcal{T}_{ik}}\,,
\end{align}
where $\mathcal{T}_\text{min}^{(i)}$ and $\mathcal{T}_\text{GM}^{(i)}$ quantifies the minimal and the average teleportation capability of the system $i$, respectively. These measures can also be extended to $N$-qubit states.

%%%%%%
\section{Relation with quantum information tasks}
\label{tasks}
Multipartite entanglement is an essential resource for quantum communication, quantum computing, quantum networks, etc. In this section we discuss the pivotal role of GME in the tasks of quantum teleportation and dense coding, followed by an introduction to some operational entanglement
measures.

\subsection{Quantum teleportation and dense coding with GME}
In Ref.~\cite{TepDenC_PhysRevLett.96.060502}, Yeo and Chua presented a protocol for faithful teleportation of an arbitrary two-qubit state using a genuine four-qubit entangled state $\ket{\chi^{00}}$, which is compared to other multipartite entangled states such as GHZ and $W$ states, then introduced a dense coding scheme using the same four-qubit state as a shared resource. Afterwards, Ref.~\cite{PhysRevA.74.032324} generalized the results of Ref.~\cite{TepDenC_PhysRevLett.96.060502} to teleport an arbitrary $N$-qubit state via genuine $N$-qubit entanglement channels.

By using a special $2N$-qubit entanglement channel $\ket{G_{1, 1, \cdots,1}}_{\mathbf{A};\mathbf{B}}$
(here $\mathbf{A};\mathbf{B}$ represents Alice's $N$ qubits and Bob's $N$ qubits of the channel respectively),
the protocol was proposed to teleport an arbitrary $N$-qubit state
\begin{equation}\label{Nq}
  \ket{\phi}_{\mathbf{a}}=\alpha_0\ket{\overline{0}}+\alpha_1\ket{\overline{1}}+\cdots+\alpha_{2^N-1}\ket{\overline{2^N-1}},
\end{equation}
where $\mathbf{a}$ are the $N$ qubits to be teleported from Alice to Bob, ${\alpha_0, \alpha_1, \cdots,\alpha_{2^N-1}}$ are complex coefficients satisfying ${|\alpha_0|^2+|\alpha_1|^2+\cdots+|\alpha_{2^N-1}|^2=1}$. And ${\ket{\overline{0}},\ket{\overline{1}}, \cdots,\ket{\overline{2^N-1}}}$ are binary forms of the states $\ket{00\cdots 0},\ket{0\cdots 01},\cdots,\ket{1\cdots 11}$.

Denote the four Bell states as
\begin{equation}
\begin{aligned}
    &\ket{\Phi_{1,2}}_{AB}=\frac{1}{\sqrt{2}}(\ket{00}_{AB}\pm\ket{11}_{AB})\,, \\
    &\ket{\Phi_{3,4}}_{AB}=\frac{1}{\sqrt{2}}(\ket{01}_{AB}\pm\ket{10}_{AB})\,.
\end{aligned}
\end{equation}
The state $\ket{G_{1, 1, \cdots,1}}_{\mathbf{A};\mathbf{B}}$ is a tensor product of $N$ Bell states $\ket{\Phi_1}_{A_1,B_1}\otimes\ket{\Phi_1}_{A_2,B_2}\otimes\cdots\otimes\ket{\Phi_1}_{A_N,B_N}$, which is expressed as
\begin{equation}\label{2NqChannel}
    \begin{aligned}
         \ket{G_{1, 1, \cdots,1}}&_{\mathbf{A};\mathbf{B}}=\frac{1}{\sqrt{2^N}}\Bigl[\ket{\overline{0}}_A\ket{\overline{0}}_B\\
         &+\cdots+\ket{\overline{2^N-1}}_A\ket{\overline{2^N-1}}_B\Bigr]\,.
    \end{aligned}
\end{equation}
Then, consider the $4^N$ generalized Bell states
\begin{equation}\label{GBS}
    \begin{aligned}                         \ket{G_{i1, i2, \cdots,iN}}&_{\mathbf{A};\mathbf{B}}=\ket{\Phi_{i1}}_{A_1,B_1}\\
    &\otimes\ket{\Phi_{i2}}_{A_2,B_2}\otimes\cdots\otimes\ket{\Phi_{iN}}_{A_N,B_N}\,,
    \end{aligned}
\end{equation}
where $i1,i2,\cdots,iN=1,2,3,4$.

Now consider the teleportation by using the maximally entangled state (MES) channel
\begin{equation}\label{MES}
  \ket{\text{MES}}_{\mathbf{A};\mathbf{B}}=\frac{1}{\sqrt{2^N}}\sum_{J=1}^{2^N}\ket{J}_A\ket{J'}_B\,,
\end{equation}
where $\ket{J}_{A}$s, $\ket{J'}_{B}$s are orthogonal bases of
the Hilbert space of Alice’s $N$ qubits and Bob’s $N$ qubits, respectively. It is noted that $\ket{\text{MES}}_{\mathbf{A};\mathbf{B}}$ may be a \emph{genuine}
multipartite entangled state, which cannot be reduced to the tensor product of $N$ Bell states in terms of $2N$ partities. Using local unitary operators applied by Alice and Bob, $\ket{\text{MES}}_{\mathbf{A};\mathbf{B}}$ connects with $\ket{G_{1, 1, \cdots,1}}_{\mathbf{A};\mathbf{B}}$ as
\begin{equation}\label{MEScGBS}
  \ket{\text{MES}}_{\mathbf{A};\mathbf{B}}=U_1(AN)U_2(BN)
  \ket{G_{1, 1, \cdots,1}}_{\mathbf{A};\mathbf{B}}\,,
\end{equation}
where unitary operators $U_1(AN)$ and $U_2(BN)$ act on $N$ qubits of Alice and Bob, respectively. Then, The joint state of the initial state of the $N$ qubits to be teleported and the entanglement
channel can be expressed as
\begin{align}\label{jointS}
\ket{\phi}&_{\mathbf{a}}\otimes  \ket{\text{MES}}_{\mathbf{A};\mathbf{B}}\nonumber\\
=&\sum_{i1, i2, \cdots,iN=1}^{4}U_1(AN)U_2(aN)\ket{G_{i1, i2,\cdots,iN}}_{\mathbf{A};\mathbf{a}}\nonumber\\
&\otimes U(i1)^{-1}U(i2)^{-1}\cdots U(iN)^{-1}\ket{\phi}_{\mathbf{B}}\,,
\end{align}
where $\ket{\phi}_{\mathbf{B}}$ is the same as the initial state of qubits $\mathbf{a}$ in Eq.~\eqref{Nq}. Here, ${U(1)=I}$, ${U(2)=\sigma_z}$, ${U(3)=\sigma_x}$, ${U(4)=i\sigma_y}$, with $\sigma_{x,y,z}$ being the three Pauli operators. If Alice performs a complete projective measurement jointly on the $2N$ qubits and gets a outcome $U_1(AN)U_2(aN) \ket{G_{i1, i2, \cdots,iN}}_{\mathbf{A};\mathbf{a}}$, then Bob will achieve an accurate replication of $\ket{\phi}_{\mathbf{a}}$ after he performs the unitary operations $U(i1)\otimes U(i2)\otimes \cdots \otimes U(iN)$ on his qubits in accordance with Alice's result of the measurement. Hence, teleportation is achieved by Alice and Bob.

It is shown that the ME channel in Eq.~\eqref{MEScGBS} is necessary and sufficient for the faithful teleportation of an $N$-qubit state together with $2N$ bits of classical communication. The GHZ states and $W$ states of $2N$ qubits, however, cannot be transferred into a maximally entangled state through local operations performed by Alice and Bob, thus they cannot be effectively utilized for faithful teleportation of an arbitrary $N$-qubit state.

The dense coding scheme using genuine four-partite entangled state $\ket{\chi^{00}}$ communicate $4$ bits of classical information by sending two particles perfectly, which is impossible with a four-partite GHZ or $W$ state \cite{TepDenC_PhysRevLett.96.060502}.
The maximally entangled five-qubit state (exhibits GME) $\ket{\psi_5}=\frac{1}{2}(\ket{001}\ket{\Phi_4}+\ket{010}\ket{\Phi_2}+\ket{100}\ket{\Phi_3}+\ket{111}\ket{\Phi_1})$ as a resource for superdense coding is discussed in Ref.~\cite{PhysRevA.77.032321}. In this case, they assume Alice owns the first three qubits and Bob has the other two of $\ket{\psi_5}$. Then Alice applies one of the sets of unitary transformations on her particle, which is the process of encoding. After having performed one of the operations described above, Alice sends off her qubits to Bob using conventional physical medium. Bob can then perform a five-partite measurement to extract the classical bits, which corresponds to decoding. It is noted that Alice and Bob have agreed on dense coding rules before encoding. The capacity of superdense coding for the state $\ket{\psi_5}$ equals to $5$, reaching the Holevo bound of maximal amount of classical information that can be encoded.

\subsection{Operational ME measures}
In Secs.~\ref{MEMea} and \ref{GMEMea}, we have reviewed various entanglement measures that are defined by axiomatic ansatz. Here we move on to the operationally meaningful measures which can quantify the value of a state for a specific protocol that requires entanglement.

\subsubsection{Distillable entanglement and entanglement cost}\label{DE and EC}
Non-maximally entangled states are insufficient for faithful teleportation as entanglement plays a crucial role in the process. However, by having a sufficient number of copies of a non-maximally entangled state, one can achieve asymptotically faithful teleportation at a certain rate. Considering a scenario where a large number $n$ of copies of a state $\rho$ are given, the question arises whether it is possible to transform $\rho^{\otimes n}$ to an output state $\sigma^{\otimes m}$ using LOCC operations, where LOCC protocols are assumed to be trace-preserving. To address this question from a physical perspective, we introduce the asymptotic limit approach, and the distance between the state resulting from the distillation protocol and the desired state is measured using the trace norm. If ${n\to \infty}$, the arbitrarily good approximation of $\sigma^{\otimes m}$ by $\sigma_m$ becomes possible. In other words, one can transform $\rho^{\otimes n}$ to an output state $\sigma_m$ that approximates $\sigma^{\otimes m}$ very well for sufficiently large $m$ with the achievable rate $r=m/n$.

The distillable entanglement is the supremum of the rates at which maximally entangled states can be obtained from an input supply of states in the form of $\rho$ over all possible distillation protocols. Denote a general trace-preserving LOCC operation by $\Lambda$, and let $\Phi(d)$ represent the density operator corresponding to the maximally entangled state vector in dimension $d$, then the distillable entanglement is defined as
\begin{equation}\label{disEnt}
  E_D(\rho)= \sup\biggl\{r: \lim_{n\to \infty}\Bigl[\inf_{\Lambda}\|\Lambda(\rho^{\otimes n})-\Phi(2^{rn})\|_1\Bigr]=0\biggr\}\,.
\end{equation}
It is noted that the state $\Phi(2^{rn})$ is local unitarily equivalent to $rn$ copies of the two-qubit maximally entangled state $\Phi(2)^{\otimes rn}$.

Now consider how many qubits we have to communicate in order to create a state $\rho$. The entanglement cost quantifies the maximal possible rate $r$ at which one can generate states that are close approximations of many copies of $\rho$ using a supply of blocks of two-qubit maximally entangled states. It provides insights into the efficiency of converting entangled resources to states that approximate the target state $\rho$, which is given by
\begin{equation}\label{Entcost}
  E_C(\rho)=\inf\biggl\{r:\lim_{n\to\infty}\Bigl[\inf_{\Lambda}\|\rho^{\otimes n}-\Lambda(\Phi(2^{rn}))\|_1\Bigr]=0\biggr\}\,.
\end{equation}
Computing $E_D(\rho)$ and $E_C(\rho)$ is a challenging task, however, these measures have significant  implications in the investigation of channel capacities, and also closely related to the entanglement of formation and relative entropy of entanglement \cite{plenio2007introduction, PhysRevA.107.012429, 10477880}.

For the multipartite scenario, the definition of distillable entanglement can be tailored to specific target states, such as GHZ states or cluster states, capturing different properties of the state being considered. Similar situations arise when defining the entanglement cost, where singlet states are commonly used as the resource, but other resources like GHZ or $W$ states can also be considered. Different experimental settings and resource availability motivate various definitions of entanglement cost and distillable entanglement. The operational point of view is crucial in understanding and interpreting these measures.

\begin{figure}
	\centering
    \begin{overpic}[width=2.5in]{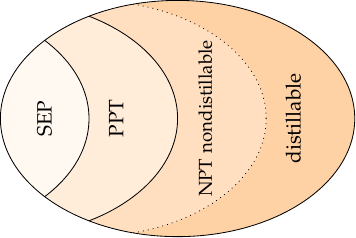}\end{overpic}
\caption{Schematic representation of the set of all states for bipartite systems, including the SEP states, PPT entangled states, hypothetical NPT nondistillable states, and distillable states.}
\label{fig2}
\end{figure}

Bound entangled states are closely related to the problem of entanglement distillation.
Generally, states that are entangled yet not distillable are termed as \emph{
bound entangled} \cite{RevModPhys.81.865, Gaida_2023}, which include positive partial transpose (PPT) entangled states. Figure~\ref{fig2} illustrates the set of all states for \emph{bipartite} systems, including separable states (SEP), PPT entangled states, \emph{hypothetical} non-PPT (NPT) nondistillable states (or NPT bound entangled states), and distillable states \cite{ PhysRevA.61.062312, PRXQuantum.3.010101}. It is clear that separable states cannot be distilled, and all PPT states are not distillable.
Notably, for $2\otimes 2$ and $2\otimes 3$ systems, it is known that all PPT states are separable and all NPT states are distillable. However, whether the hypothetical NPT nondistillable states exist remains an important and open question.

\subsubsection{Multipartite distillable key}\label{DistillableKey}
The distillable (secret) key quantifies the asymptotic rate at which Alice and Bob may distill secret classical bits from many copies of a shared quantum state.
Rather than focusing on distilling singlets, one can explore the distillation of private states, which encompasses a broader class of states. Let's start with the $m$-partite private state, or $m$-partite \emph{pdit} \cite{PhysRevA.74.010302}
\begin{equation}\label{pdit}
  \Gamma^{(d)}=\sum_{i,j=0}^{d-1}\frac{1}{d}\ket{ii\cdots i}\!\bra{jj\cdots j}\otimes U_i \rho_{A_1^{\prime}, A_2^{\prime}, \cdots,A_m^{\prime}}U_j^{\dagger}\,,
\end{equation}
where $\rho_{A_1^{\prime}, A_2^{\prime}, \cdots,A_m^{\prime}}$ denotes a state on the Hilbert space $\cH_{A_1^{\prime}, A_2^{\prime}, \cdots,A_m^{\prime}}$, and $U_i (i=0, 1, \cdots,d-1)$ are some unitary operations. Thus the state $\Gamma^{(d)}$ is defined on the Hilbert space $(\cH_{A_1}\otimes \cH_{A_2}\otimes\cdots \otimes \cH_{A_m})\otimes (\cH_{A_1^{\prime}}\otimes\cH_{A_2^{\prime}}\otimes \cdots \otimes \cH_{A_m^{\prime}})$. The system $A_1, A_2, \cdots,A_m$ giving secure bits of key is called the \emph{key part} of the private state, and the system $A_1^{\prime}, A_2^{\prime}, \cdots,A_m^{\prime}$ defends the key in system $A_1, A_2, \cdots,A_m$ from eavesdropper is called the \emph{shield part}.

For a given multipartite state $\rho_{A_1, A_2, \cdots,A_m}$ acting on ${\cH_{A_1}\otimes\cH_{A_2}\otimes \cdots \otimes\cH_{A_m}}$, consider a sequence $\Lambda_n$ of LOCC operations such that ${\Lambda_n\bigl(\rho_{A_1, A_2, \cdots,A_m}^{\otimes n}\bigr)=\sigma^{(n)}}$, where $n$ is the number of copies of the state. A set of operations ${\cP=\cup_{n=1}^{\infty}\{\Lambda_n\}}$ is called
a pdit distillation protocol of state $\rho_{A_1, A_2, \cdots,A_m}$ if the condition
\begin{equation}\label{pditDisPro}
  \lim_{n\to\infty}\|\sigma^{(n)}-\Gamma^{(d_n)}\|_1=0
\end{equation}
holds.
Here $\Gamma^{(d_n)}$ is a multipartite pdit whose key part is of dimension $d_n\times d_n$.
For a protocol $\cP$, its rate is given by taking the limit superior of $(\log_2 {d_n})/n$, i.e.,
\begin{equation}\label{proRate}
  \cR(\cP)=\limsup_{n\to\infty}\frac{\log_2 {d_n}}{n}\,.
\end{equation}
Then the distillable key of state $\rho_{A_1, A_2, \cdots,A_m}$ is given by \cite{PhysRevA.80.042307}
\begin{equation}\label{disKey}
  K_D^{(m)}(\rho_{A_1, A_2, \cdots,A_m})=\sup_{\cP}\cR(\cP)\,.
\end{equation}
For the $N$-partite
quantum state $\rho$, the normalized multipartite squashed entanglement is an upper bound on the distillable key \cite{5075874}
\begin{equation}\label{dkupperbound}
  K_D^{(m)}(\rho)\le\frac{1}{m}E_{sq}^q(\rho)\,.
\end{equation}
Another upper bound of the distillable key is the regularized version of the relative entropy \cite{PhysRevA.80.042307}. For a tighter upper bound, one can consider the regularized relative
entropy with respect to biseparable states \cite{PhysRevX.11.041016}.

When considering the distribution of secret keys, the multipartite private states, which are the output of a protocol that distills secret key among trusted parties, are necessarily genuine multipartite entangled \cite{PhysRevX.11.041016}. Besides, the distillation of private states is the goal of quantum key repeater, a device that extends the distance of quantum key distribution (QKD) while being capable of tolerating higher noise levels and beyond the limitations of entanglement distillation \cite{bauml2015limitations, PhysRevLett.119.220506}.

\subsubsection{Accessible entanglement and source entanglement}
If a quantum state $\rho$ can be transformed to another quantum state $\sigma$ by LOCC, then $\rho$ is considered to be at least as useful as $\sigma$. This is because any application or protocol that can be achieved using $\sigma$ can also be achieved using $\rho$, but the reverse may not be true. If the parties are provided with $\rho$, they can transform it to $\sigma$ without incurring any additional cost and proceed with the desired protocol. The study of LOCC transformations allows us to determine the relative usefulness of different quantum states and forms the basis for quantifying entanglement. And LOCC corresponds to those operations which can be implemented without depleting the existing entanglement resources, thus entanglement measures must be quantities that are nonincreasing under these LOCC transformations.

In order to further understand multipatite entanglement, entanglement measures with an operational meaning for arbitrary multipartite states (pure or mixed) of any dimension are introduced \cite{PhysRevLett.115.150502}. These measures, known as \emph{accessible} and \emph{source} entanglement, can be computed once the possible LOCC transformations are characterized. If there exists a deterministic LOCC protocol that can transform state $\ket{\psi}$ to state $\ket{\phi}$, we say that state $\ket{\psi}$ can \emph{reach} state $\ket{\phi}$, and state $\ket{\phi}$ is \emph{accessible} from state $\ket{\psi}$. For a given state $\ket{\psi}$, the accessible set $M_a(\ket{\psi})$ is defined as the set of states that are reachable via LOCC from $\ket{\psi}$. And the source set, $M_s(\ket{\psi})$ is the set of states that can reach $\ket{\psi}$. Then consider $\mu$ as an arbitrary measure in the set of local unitarily equivalent classes, the corresponding accessible volume which measures
the amount of states that can be accessed by $\ket{\psi}$, and the source volume which measures the amount of
states that can be used to reach the state are ${V_a(\ket{\psi})=\mu [M_a(\psi)]}$ and ${V_s(\ket{\psi})=\mu [M_s(\psi)]}$, respectively. The intuition is that the larger the accessible volume is, the more valuable the state is, while the larger the source volume is, the less powerful the state is. Thus, the accessible entanglement and the source
entanglement are given by
\begin{equation}\label{accEnt&souEnt}
    E_a(\ket{\psi})=\frac{V_a(\ket{\psi})}{V_a^{\sup}}\,,\,\,\, E_s(\ket{\psi})=1-\frac{V_s(\ket{\psi})}{V_s^{\sup}}\,,
\end{equation}
where $V_a^{\sup}$ and $V_s^{\sup}$ represent the maximally accessible volume and the maximally source volume according to the measure $\mu$, respectively.
It is worth noting that the operational properties of these measures make it easy to prove that they are nonincreasing under LOCC. Also, ${M_s(\ket{\psi})=\emptyset}$ iff the state is in the MES because the maximally entangled states are the only states that cannot be reached by any other state \cite{PhysRevLett.111.110502}. For example, the $W$ state achieves the maximum values of both ${E_a(\ket{W})=1}$ and ${E_s(\ket{W})=1}$. This highlights the exceptional nature of the $W$ state, making it the most valuable state within the $W$ class.

\subsubsection{Concentratable entanglement}
As a computable and operationally meaningful family of entanglement measures, the concentratable entanglement is introduced in Ref.~\cite{PhysRevLett.127.140501}. Consider an $N$-qubit pure quantum state $\ket{\phi}$, $\cS=\{1,2,\cdots,N\}$ as
the set of labels for each qubit and its power set $\cP(\cS)$, for any set of qubit labels ${s\in \cP(\cS)\!\setminus\!\{\emptyset\}}$, the concentratable entanglement is defined as
\begin{equation}\label{ConEnt}
  \cC_C(s)=1-\frac{1}{2^{c(s)}}\sum_{\alpha\in\cP(s)}\Tr(\rho_{\alpha}^2)\,,
\end{equation}
where $c(s)$ is the cardinality, and $\rho_{\alpha}$ represents the joint reduced state. The purity $\Tr(\rho_{\alpha}^2)$ can be computed via an overlap test, for instance to employ the $N$-qubit parallelized {\sc swap} test to compute the state overlap ${|\braket{\phi_1}{\phi_2}|^2}$.
For the example of $N$-qubit $\ket{W}$ and $\ket{\textrm{GHZ}}$ state, one finds that $\cC_C^{\ket{\textrm{GHZ}}}(s)>\cC_C^{\ket{W}}(s)$ and the concentratable entanglement detects more ME in $\ket{\textrm{GHZ}}$ than in $\ket{W}$.

The $N$-tangle, concurrence, and linear entropy of entanglement can be seen as special cases of concentratable entanglement. For two-qubit systems, $\cC_C(s)=\tau_{(2)}/4=\cC^2/4$ for all $s\in\cP(\cS)$, where $\tau_{(N)}=|\braket{\phi}{\tilde{\phi}}|^2$ with $\ket{\tilde{\phi}}=\sigma_y^{\otimes N}\ket{\phi^\ast}$ and $\cC$ is the concurrence. Concentratable entanglement quantifies the probability that at least one {\sc swap} test fails when $N$ of them are applied
in parallel across two copies of $\ket{\phi}$. It is operationally interpreted as the probability of producing Bell pairs in the parallelized {\sc swap} test.
The controlled-{\sc swap} (c-{\sc swap}) operation is an experimentally accessible tool and a fundamental component in quantum communication protocols for measuring and witnessing entanglement.

%%%%%%
\section{Summary and Outlook}
\label{Conclusions}
In this review, we have surveyed various ME measures with an emphasis on the genuine and operational ME measures.
The intention behind this review is to provide valuable insights and serve as a helpful resource for future research in the field of quantum information processing.
In addition, we hope that this review would inspire and guide researchers in their endeavors to further develop novel approaches for characterizing ME.

In spite of continuous
progress, the current status of entanglement measure theory is still marked by a number of outstanding open problems, some of which include
\begin{enumerate}[1)]
 \item The quest for an effective and universally applicable measure to quantify GME in systems involving more than three parties remains as a significant challenge.
 \item While theoretical entanglement measures provide valuable insights of quantum entanglement, operational measures are designed to be practically applicable in quantum information processing tasks. Thus, the exploration and advancement of new operational measures for ME is essential.
 \item  Extending entanglement measures to mixed-state systems is also a meaningful direction. Real-world quantum state is mixed due to unavoidable interactions with the environment. Thus, it is important to derive computable bound for existing entanglement measures or design efficient entanglement measures for multipartite mixed states.
\end{enumerate}

In summary, we believe that the research on ME measures is ongoing and is likely to have a crucial impact on other tasks in the era of quantum information science.

%%%%%%
\section{Declaration of Competing Interest}
The authors declare that they have no conflicts of interest in this work.

%%%%%%
\acknowledgments
This work was supported by the National Natural Science Foundation of China (Grants No.~92265115 and No.~12175014) and the National Key R\&D Program of China (Grant No.~2022YFA1404900).

%%%%%%
% \bibliographystyle{apsrev4-1}
%\bibliography{Refs}

%apsrev4-2.bst 2019-01-14 (MD) hand-edited version of apsrev4-1.bst
%Control: key (0)
%Control: author (8) initials jnrlst
%Control: editor formatted (1) identically to author
%Control: production of article title (0) allowed
%Control: page (0) single
%Control: year (1) truncated
%Control: production of eprint (0) enabled

%

\end{document}